\documentclass[12pt]{article}
\usepackage{graphicx}
\usepackage{amssymb}
\usepackage{amsmath}

\newcommand{\bear}{\begin{array}}  
\newcommand {\eear}{\end{array}}
\newcommand{\bea}{\begin{eqnarray}}   
\newcommand{\eea}{\end{eqnarray}}
\newcommand{\beq}{\begin{equation}}   
\newcommand{\eeq}{\end{equation}}
\newcommand{\bef}{\begin{figure}}  \newcommand 
{\eef}{\end{figure}}
\newcommand{\bec}{\begin{center}}  \newcommand 
{\eec}{\end{center}}

\begin{document}

\begin{titlepage}

\begin{flushright}
ICRR-Report-580-2010-13\\
IPMU 11-0015 \\
KEK-TH-1442\\
UT-11-05\\
\end{flushright}

\vskip 1.35cm


\begin{center}

{\large \bf
Cosmological constraints on dark matter models with
velocity-dependent annihilation cross section
}

\vskip 1.2cm

Junji Hisano$^{a,b,c}$,
Masahiro Kawasaki$^{b,c}$,
Kazunori Kohri$^{d,e,f}$,
Takeo Moroi$^{g,c}$,\\
Kazunori Nakayama$^{d}$ and
Toyokazu Sekiguchi$^{b}$

\vskip 0.4cm
{\it $^a$Department of Physics, Nagoya University, Nagoya 464-8602, Japan}\\
{ \it $^b$Institute for Cosmic Ray Research,
University of Tokyo, Kashiwa 277-8582, Japan}\\
{\it $^c$Institute for the Physics and Mathematics of the Universe,
University of Tokyo, Kashiwa 277-8568, Japan}\\
{\it $^d$Cosmophysics Group, Theory Center, IPNS, KEK, Tsukuba, 305-0801, Japan}\\
{\it $^e$Department of Particle and Nuclear Physics, The Graduate
University for Advanced Studies, Tsukuba, 305-0801, Japan}\\
{\it $^f$Department of Physics, Tohoku University, Sendai 980-8578, Japan }\\
{\it $^g$Department of Physics, University of Tokyo, Bunkyo-ku, Tokyo 113-0033, Japan}

\date{\today}

\begin{abstract} 
  We derive cosmological constraints on the annihilation cross section of
  dark matter with velocity-dependent structure, motivated by
  annihilating dark matter models through Sommerfeld or Breit-Wigner
  enhancement mechanisms.  In models with annihilation cross section
  increasing with decreasing dark matter velocity, big-bang
  nucleosynthesis and cosmic microwave background give stringent
  constraints.
\end{abstract}



\end{center}
\end{titlepage}

\section{Introduction}

In a weakly-interacting massive particle dark matter (WIMP DM)
scenario, a DM particle with mass of $\mathcal{O}(100)$~GeV --
$\mathcal{O}(1)$~TeV should have an (thermally-averaged) annihilation cross section of
$\langle\sigma v\rangle\simeq 3\times 10^{-26}~{\rm cm^3/s}$ in order
to reproduce the observed DM abundance due to the thermal production.
On the other hand, recently reported excesses of cosmic-ray
positron~\cite{Adriani:2008zr} and electron
fluxes~\cite{:2008zz,Abdo:2009zk,Collaboration:2008aaa} may be
interpreted as signatures of annihilating dark matter with fairly
large annihilation cross section of order of $~10^{-23}$ --
$10^{-22}~{\rm cm^3/s}$ depending on DM mass $m$, which is typically
three orders of magnitude larger than the canonical value quoted
above, although constraints from other observations, such as
gamma-rays~\cite{Profumo:2009uf,Abdo:2010ex,Zaharijas:2010ca} and
neutrinos~\cite{Hisano:2008ah,Liu:2008ci,Spolyar:2009kx,Abbasi:2011eq}
are also stringent and might have already excluded some parameter
regions.

One way to achieve the ``boost factor'' of $\mathcal{O}(10^3)$ is to
make the DM annihilation cross section velocity-dependent.  In this
case the annihilation cross section in the early Universe is not same
as that  in the Galaxy or elsewhere, simply because typical velocity
of the DM particle varies from place to place.  Hence it is in
principle possible that the DM has canonical annihilation cross
section at the freezeout epoch in the early Universe reproducing the
DM abundance observed by Wilkinson Microwave Anisotropy Probe (WMAP),
while explaining the cosmic-ray
positron/electron excesses.  A common mechanism would be Sommerfeld
enhancement of annihilation cross section~\cite{Hisano:2003ec,Hisano:2005ec,
ArkaniHamed:2008qn}.  If a DM interacts with a light particle through
which it annihilates, non-perturbative effects enhance the
annihilation cross section.  The cross section is enhanced by inverse
of the DM velocity, $v^{-1}$ or $v^{-2}$, in this class of models.  In
the Breit-Wigner enhancement scenario on the other hand, DM
annihilates through $S$-channel resonance where a particle in the
intermediate state has a mass close to two times DM
mass~\cite{Feldman:2008xs,Ibe:2008ye,MarchRussell:2008tu}.  In this case the DM cross section can scale
as $v^{-4}$ at an earlier time or $v^{-2}$ at a later time.

In these models the annihilation cross section increases as the
temperature decreases in the early Universe, and hence DM continues to
inject high energy particles through the cosmic history.  Therefore,
it is quite non-trivial whether these models satisfy constraints from
big-bang nucleosynthesis (BBN) and cosmic microwave background (CMB).
In the case of velocity-independent annihilation cross section, bounds
from BBN~\cite{Reno:1987qw,Frieman:1989fx,
  Jedamzik:2004ip,Hisano:2008ti,Hisano:2009rc} and
CMB~\cite{Padmanabhan:2005es,Galli:2009zc,Kanzaki:2009hf} were derived
in previous works.  In this paper, we extend the analysis to the
velocity-dependent annihilation cross section and derive general upper
bound on the annihilation cross section.

This paper is organized as follows.  In Sec.~\ref{sec:DM} a simple
prescription for treating the velocity-dependence of DM annihilation
cross section is described.  In Sec.~\ref{sec:constraints} we present
constraints from BBN and CMB and give implications on DM models.
Sec.~\ref{sec:conc} is devoted to conclusions and discussion.


\section{Dark matter with velocity-dependent cross section} 
\label{sec:DM}

\subsection{Models of velocity-dependent annihilation cross section}

Below we briefly give examples of DM with velocity-dependent
annihilation cross section.  After that we will explain our unified
treatment for describing the cosmological effects from DM annihilation
with velocity-dependent annihilation cross section.

\subsubsection{Sommerfeld enhancement}

A DM particle $\chi$ is assumed to have an interaction with $\phi$,
which may be a scalar or gauge boson with coupling constant
$\alpha_\chi$, whose mass is much lighter than the DM mass: $m_\phi
\ll m$.  Let us consider the DM annihilation process mediated by
$\phi$ exchanges.  If the mass of $\phi$ is sufficiently small, the
$\phi$-mediated interaction can be regarded as a long-range force and
such an annihilation cross section receives an enhancement $S$
compared with tree-level perturbative expression~\cite{Sommerfeld},
\begin{equation}
        S = \frac{\pi \alpha_\chi / v}{1-e^{-\pi \alpha_\chi/v}},
\end{equation}
where $v$ is the initial DM velocity in the center of mass frame.
Thus the DM annihilation cross section is proportional to $1/v$ for $v \ll \alpha_\chi$.
This $1/v$ enhancement saturates at $v \sim m_\phi/m$.
There is another interesting effect caused by the bound state
formation, which resonantly enhances the DM annihilation rate for some
specific DM mass ~\cite{Hisano:2003ec,ArkaniHamed:2008qn}.  It is
known that the enhancement is proportional to $v^{-2}$ near the
zero-energy resonance, and this $v^{-2}$ behavior also saturates due to the finite
width of the bound state. 
%

\subsubsection{Breit-Wigner enhancement}

In the Breit-Wigner enhancement scenario,
DM particles annihilate through $S$-channel particle exchange ($\phi$),
where the mass of $\phi$, $m_\phi$, is close to $2m$.
The square amplitude of this $S$-channel process is proportional to
\begin{equation}
        |\mathcal M|^2 \propto \frac{1}{(v^2 + \delta)^2 + \gamma^2},
\end{equation}
where $\delta$ and $\gamma$ are defined as $m_\phi^2 = 4m^2(1-\delta)$
and $\gamma=\Gamma_\phi/m_\phi$ ($\Gamma_\phi$ is the decay width of
$\phi$), respectively.  If $\delta$ and $\gamma$ are much
smaller than unity, we have $\langle \sigma v \rangle \propto v^{-4}$ in the limit
$v^2 \gg \mbox{max}[\delta,\gamma]$.  At smaller velocity, it becomes proportional to
$v^{-2}$.  Finally, in  sufficiently small $v$  the cross section saturates at
a constant value.  

%

\subsection{Energy injection from DM annihilation}

Some DM models have velocity-dependent annihilation cross section as
described above.  In order to treat the effects of
velocity-dependence, we phenomenologically parametrize the
annihilation cross section as
\begin{equation}
  \langle \sigma v\rangle = 
  \frac{\langle\sigma v\rangle_0}{\epsilon + \left( v/v_0 \right)^n},   
  \label{sigmav}
\end{equation}
where $\langle \sigma v\rangle_0$ is a constant, $v$ is the
(thermal-averaged) velocity of DM particle, and $v_0$ is the velocity at
the freezeout of DM annihilation in the velocity-independent case.
Typically, the freezeout temperature is given by $T_{\rm fo}\sim
  m/25$~\cite{Kolb:1990}, which gives $v_{0} \sim \sqrt{3}/5 \sim
  0.3$; in our numerical calculations, we take $v_0 = \sqrt{3}/5$
  independently of $n$ although the freezeout epoch may deviate from
  $T_{\rm fo}$ given above. In Eq.\ \eqref{sigmav}, $\epsilon$ is a
dimensionless parameter which determines the cutoff below which the
velocity-dependence disappears.  Since we are interested in the case
that $\epsilon\ll 1$, we recover $\langle \sigma v\rangle \simeq
\langle \sigma v\rangle_0$ in the limit $v\to v_0$.

The power law index $n$ and the cutoff parameter $\epsilon$ depend on
models. The Sommerfeld enhancement predicts $n=1$ and $\epsilon \simeq
m_\phi/m$, while it also predicts $n=2$ in the zero-energy resonance
region.  In the Breit-Wigner enhancement, the DM annihilation cross
section reduces to the form (\ref{sigmav}) with $n=4$ and $\epsilon =
[ \delta^2 + \gamma^2 ]/v_0^4$ in the limit $v^2 \gg
\mbox{max}[\delta,\gamma]$.  In another limit $v^2 \ll
\mbox{max}[\delta,\gamma]$, it is of the form with $n=2$ and $\epsilon
= [ \delta^2 + \gamma^2 ]/(2\delta v_0^2)$, after rescaling
$\langle\sigma v\rangle_0 \to \langle\sigma v\rangle_0 v_0^2/2\delta$.
When the annihilation cross section with $n=4$, the DM annihilation 
cross section is large enough to reduce the DM number
density significantly even below the freezeout temperature. Thus, we 
consider the cases of $n=1$ and $2$ in the following analysis, and 
it would give conservative bounds on the Breit-Wigner enhancement.


With the annihilation cross section being given, the annihilation term
in the Boltzmann equation, which governs the evolution of the number
density of DM $n_{\rm DM}$, is given by
\begin{eqnarray}
  \left[ \frac{d n_{\rm DM}}{dt} \right]_{\rm ann} = 
  - n_{\rm DM}^2 \langle \sigma v\rangle.
\end{eqnarray}
In deriving constraints from BBN and CMB, spectra of injected energy
per unit time are needed for all the daughter particles. For the
particle species $i$, such a quantity is given by
\begin{eqnarray}
  \left[ \frac{d f_i (E) }{dt} \right]_{\rm ann} = 
  \frac12 n_{\rm DM}^2 \langle \sigma v\rangle 
  \frac{d N_i}{dE},
\end{eqnarray}
where $dN_i/dE$ is the energy spectrum of $i$ from the pair
annihilation of DM.  The energy spectra of decay products depend on
the property of DM; for a given decay process, we calculate
$dN_i/dE$ by using PYTHIA package~\cite{Sjostrand:2006za}.

\begin{figure}
 \begin{center}
   \includegraphics[width=0.6\linewidth]{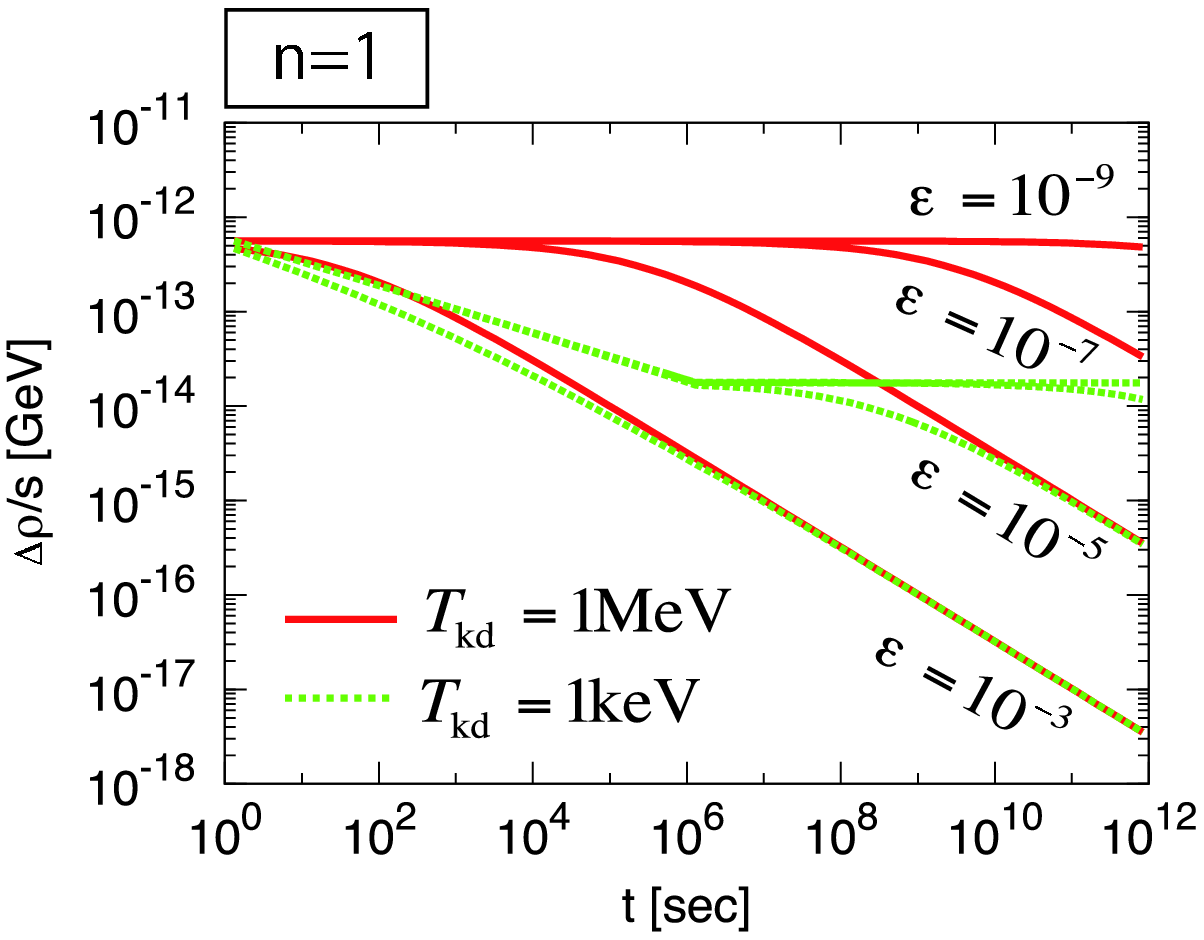}
   \vskip 1cm
   \includegraphics[width=0.6\linewidth]{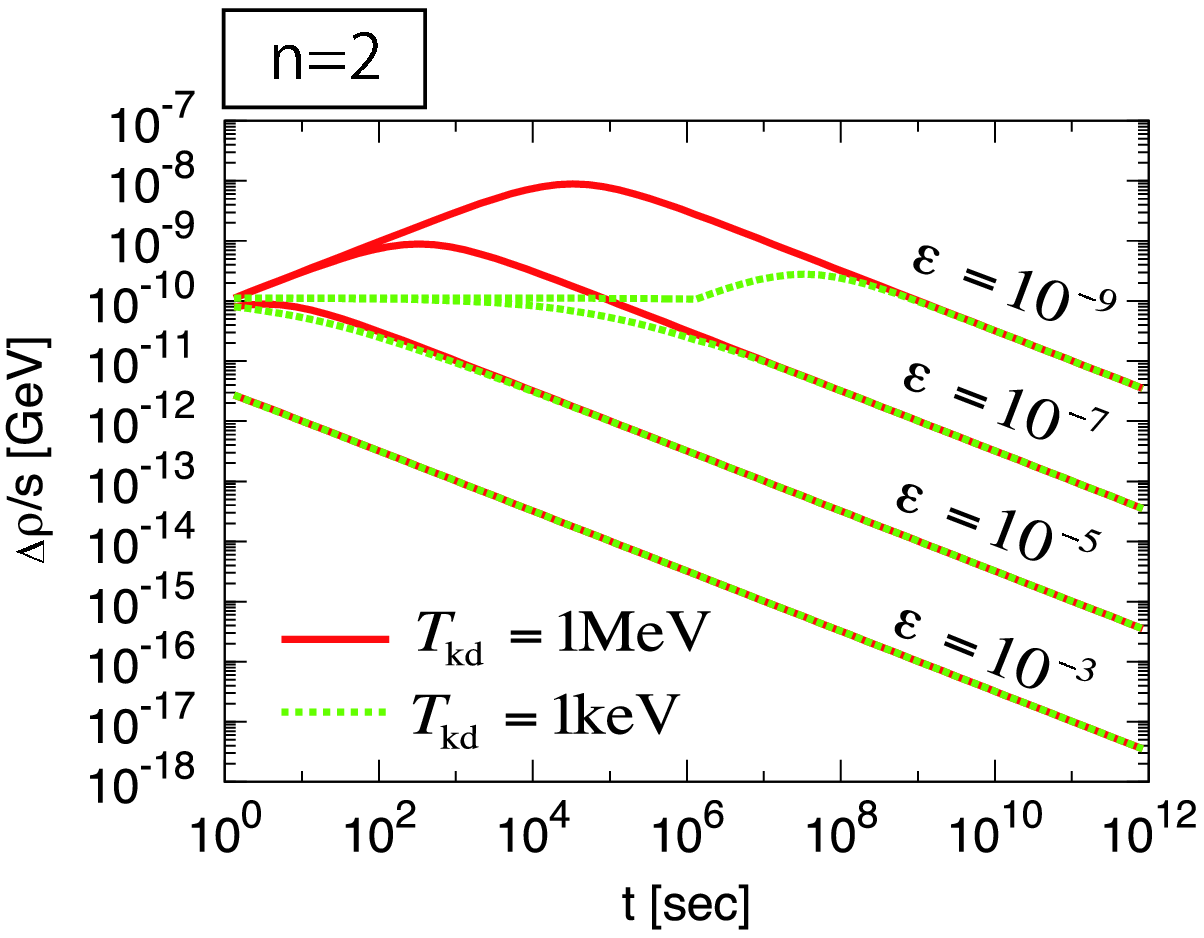} 
   \caption{ Energy injection from DM annihilation per entropy density
     per Hubble time as a function of time for $n=1$ (top panel) and
     $n=2$ (bottom panel).  Red solid lines correspond to $T_{\rm
       kd}=1$~MeV and green dashed lines correspond to $T_{\rm
       kd}=1$~keV, for $\epsilon=10^{-3},10^{-5},10^{-7},10^{-9}$ from
     bottom to top.  We have taken $\langle\sigma v\rangle_0=3\times
     10^{-26}~{\rm cm^3/s}$ and $m=1$~TeV.  }
   \label{fig:rhos}
 \end{center}
\end{figure}

 In order for a qualitative understanding of the effects of
  velocity-dependent cross section, it is instructive to consider the
  total energy injection $\Delta\rho$ in typical cosmic time, which is
  $\sim H^{-1}$, with $H$ being the expansion rate of the universe.
  For this purpose, let us define
\begin{eqnarray}
  \frac{\Delta \rho}{s}
  \equiv
  \frac{1}{2}
  \frac{E_{\rm vis}n_{\rm DM}^2 \langle \sigma v\rangle H^{-1}}{s},
\end{eqnarray}
where $E_{\rm vis}$ is the total release of visible energy in one
pair-annihilation process of DM, and $s$ is the entropy density.
Because we consider the case that the cosmic expansion is
  (almost) unaffected by the DM annihilation, the quantity $\Delta
  \rho/s$ is approximately proportional to the amount of injected
  energy in a comoving volume per Hubble time. Numerically, we
obtain
\begin{eqnarray}
  \frac{\Delta \rho}{s} \simeq
  2.9\times 10^{-18}~{\rm GeV}
  \left( \frac{T}{1~{\rm keV}} \right)
  \left( \frac{E_{\rm vis}}{2m} \right)
  \left( \frac{1~{\rm TeV}}{m} \right)
  \left( \frac{\langle \sigma v\rangle_0}{1~{\rm pb}} \right)
  {\rm min}\left[ \frac{1}{\epsilon}, 
    R_{e}
  \right],   
  \label{deltarho}
\end{eqnarray}
where $T$ is the cosmic temperature.  In addition, for the convenience
of the following discussion, we have introduced the enhancement factor
\begin{eqnarray}
    \label{eq:Renhance}
    R_{\rm e} \equiv \left(
      \displaystyle{
      \frac{v_{0}}{v}
      }
    \right)^{n}.
\end{eqnarray}
We have set the present DM energy density to be consistent with the
WMAP observation, $\Omega_c h^2 \simeq 0.11$~\cite{Komatsu:2010fb}.\footnote
{ In the standard thermal relic scenario, $\langle\sigma v\rangle_0$
  is fixed once we fix the DM abundance, though it deviates from the
  canonical value $(\sim 3\times 10^{-26}~{\rm cm^3/s})$ due to the
  velocity dependence around the DM freezeout epoch, especially in the
  case of $n=2$~\cite{Hisano:2006nn,Dent:2009bv}.  In the following we
  derive upper bound on $\langle\sigma v\rangle_0$ with fixed DM
  abundance.  }
In the case of velocity-independent cross section (where
$\langle\sigma v\rangle = \langle\sigma v\rangle_0$), it is evident
that the energy injection per comoving volume decreases as $T$
decreases.  This is natural since the DM number density decreases as
the Universe expands.  In the case of the velocity-dependent cross
section, however, some non-trivial features appear.  First note that
the velocity is estimated as
\begin{equation}
\begin{split}
        &\frac{v}{v_{0}} = \sqrt{\frac{25T}{m}} \propto
        T^{1/2} 
        \qquad \quad {\rm for} \qquad T>T_{\rm kd},\\
        &\frac{v}{v_{0}} = \sqrt{\frac{25T_{\rm kd}}{m}}
        \frac{T}{T_{\rm kd}} \propto T 
        \qquad {\rm for} \qquad T<T_{\rm kd},
\end{split}
\end{equation}
where $T_{\rm kd}$ denotes the temperature at the {\it kinetic
  decoupling}.\footnote
{Precisely speaking, there appears a dependence on the relativistic
  degrees of freedom $g_{*s}$ for $T<T_{\rm kd}$, as $v \propto
  g_{*s}(T)^{1/3}T$. We have taken into account this
    correction.
}
Below this temperature, a DM particle cannot maintain kinetic
equilibrium with thermal plasma, and hence it propagates freely and
loses its momentum only adiabatically by the Hubble expansion.  Notice
that typical kinetic decoupling temperature for WIMP DM is much
smaller than the freezeout temperature (we may say $T_{\rm kd} \sim$
keV -- MeV for WIMP DM candidates)
\cite{Kawasaki:1995cy,Hisano:2000dz,Profumo:2006bv,Bringmann:2006mu}.
Thus it is found that, for $T<T_{\rm kd}$ and $n\geq 1$, the energy
injection (\ref{deltarho}) is constant, or increases as $T$ decreases
as long as the velocity dependence is not saturated.  In Fig.\
\ref{fig:rhos}, we plot $\Delta\rho/s$ as a function of time for $n=1$
(top) and $n=2$ (bottom).  Red solid lines correspond to $T_{\rm
  kd}=1$~MeV, and green dashed lines correspond to $T_{\rm kd}=1$~keV,
for $\epsilon=10^{-3}$ -- $10^{-9}$ from bottom to top.  $T_{\rm
  kd}=1$~keV approximately corresponds to a lower bound on the
temperature of kinetic decoupling in order not to suppress the density
fluctuation for a formation of the Lyman-$\alpha$ clouds~\cite{Loeb:2005pm}. We have
taken $\langle\sigma v\rangle_0=3\times 10^{-26}~{\rm cm^3/s}$,
$m=1$~TeV and $E_{\rm vis}=2m$.  Then, even at $T\sim$ 0.1 MeV, we get
an enhancement factor of the order of $R_{\rm e} \sim 10^{3} $ $(\sim
10^{6} )$ with $n=1$ ($n=2$). This implies that constraints become
stronger than the case of usual DM without velocity dependence.



In the following we perform detailed calculations of the effects on
BBN and CMB, and derive constraints on the cross section with various
choice of $\epsilon$ and $T_{\rm kd}$.


\section{Constraints from BBN and CMB} \label{sec:constraints}

\subsection{Constraints from BBN} \label{sec:BBN}

\subsubsection{Basic picture} \label{subsubsec:picture}

It has been known that injection of high-energy particles which are
emitted through the annihilation of long-lived massive particles
during/after the big-bang nucleosynthesis epoch (at a cosmic time
$t=10^{-2}$ -- $10^{12}$~sec) significantly changes the light element
abundances~\cite{Reno:1987qw,Frieman:1989fx,Jedamzik:2004ip,Hisano:2008ti,
  Hisano:2009rc,Jedamzik:2009uy}. However, the effect of the injection
highly depends on what particles are injected. We discuss two
possibilities: (i) injection of electromagnetic particles and (ii)
injection of hadronic particles in this section.

The injection of high-energy electromagnetic particles such as photon
and electron induces the electromagnetic cascade, which produces a lot
of energetic photons. Those photons destroy the background $^{4}$He
and produce lighter elements such as deuterium (D), tritium (T),
$^{3}$He, and heavier elements such as $^{6}$Li nonthermally at $t
\gtrsim 10^{6}$ sec. In particular there is a striking feature that
the $^{3}$He to D ratio ($^{3}$He/D) tends to increase. By comparing
to the observed value of $^{3}$He/D, this gives us the most
stringent bound on the annihilation cross section in case of the
injection of electromagnetic particles~\cite{Hisano:2009rc}. This
reaction occurs at the cosmic temperature of $T \sim 10^{-4}$ MeV. It
is notable that this constraint from BBN~\cite{BBN} is stronger than
that on the $\mu$- or $y$-distortion from the Planck distribution of
CMB~\cite{Zavala:2009mi}.

On the other hand, the injection scenario of high-energy hadrons such
as pion, proton ($p$), neutron ($n$) and their antiparticles might be
more complicated, but has been understood in detail~\cite{BBN,Jedamzik:2006xz}. The
emitted high-energy neutron and proton destroy the background $^{4}$He
and produce D, T, $^{3}$He or $^{6}$Li. The charged pions, $n\bar{n}$
and $p\bar{p}$ pairs induce an extra-ordinal interconversion between
the background proton and neutron, which makes the neutron to proton
ratio ($n/p$) increase. Then this mechanism produces more $^{4}$He.
In terms of the annihilating dark matter, the overproduction of D
or the increase of $^3{\rm He}$ to deuterium ratio ($^3{\rm
    He}/{\rm D}$) gives us the most stringent constraint on the
annihilation cross section~\cite{Hisano:2008ti,Hisano:2009rc}.

In the following, we perform a detailed calculation of the
  light-element abundances; to take account of the injection of
  hadronic and electromagnetic particles, we follow the procedure
  given in \cite{BBN}.  Then, comparing the theoretical prediction with
  the updated observational constraints, we derive precise upper
  bounds on the annihilation cross section as a function of the DM
  mass.

\subsubsection{Observational light element abundances} \label{subsubsec:obs}

Next we discuss observational limits on D/H and $^{3}$He/D which are
adopted in this study.  In the previous work~\cite{Hisano:2009rc}, it
was shown that these elements give us more stringent constraints than
the others.
The recent observation of the metal-poor  
QSO absorption line system QSO~Q0913$+$072, together with 
the six previous measurements, leads to 
value of the primordial deuterium abundance with a sizable
dispersion~\cite{Pettini:2008mq},
\begin{eqnarray}
    \label{eq:Dobs}
    {\rm (n_{\rm D}/n_{\rm H})}_{\rm p} = 
    (2.82 \pm 0.20) \times 10^{-5}.
\end{eqnarray}
Compared with data adopted in the previous analyses 
in Refs.~\cite{Hisano:2008ti,Hisano:2009rc}, the error of ${(n_{\rm
    D}/n_{\rm H})}_{\rm p}$ has been reduced by about $20\ \%$.

We adopt an  upper limit on $n_{\rm ^3He}/n_{\rm D}$ which is recently
observed in protosolar clouds~\cite{GG03},
\begin{eqnarray}
    \label{eq:He3D}
    (n_{\rm ^3He}/n_{\rm D})_{\rm p} < 0.83+0.27.
\end{eqnarray}
This  value was also used in Ref.~\cite{Hisano:2009rc}.

\subsubsection{Constraints on electromagnetic particle injection} 
\label{subsubsec:elemag}

\begin{figure}
 \begin{center}
   \includegraphics[width=0.6\linewidth]{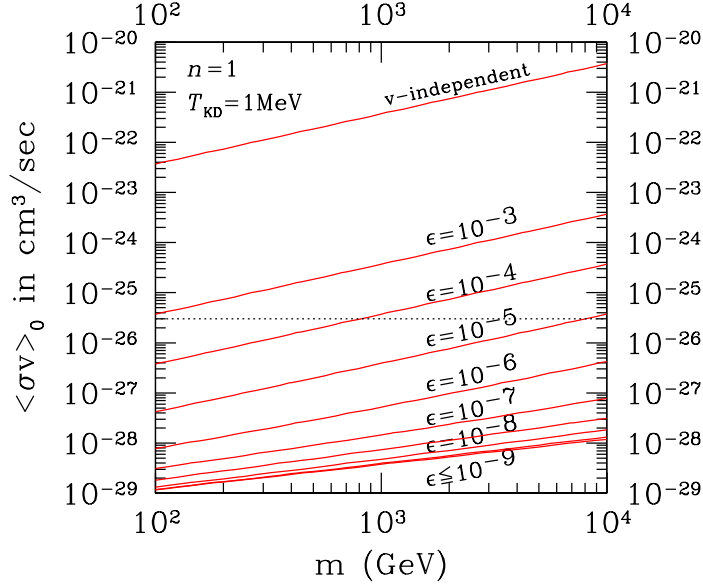}
   \includegraphics[width=0.6\linewidth]{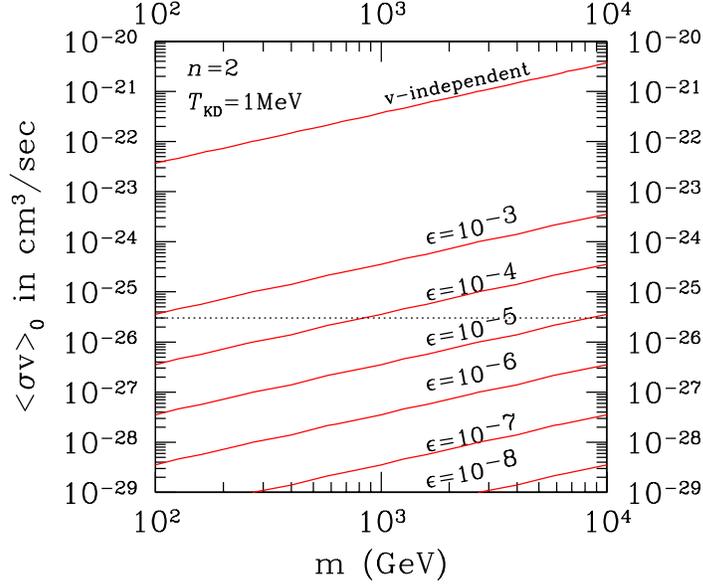} 
   \caption{ Upper bound on the annihilation cross section obtained
     from the observational $^{3}$He/D limit with $n=1$ (top) and
     $n=2$ (bottom) for various values of $\epsilon =
     10^{-10}$--$10^{-3}$.
     Here DM is assumed to annihilate purely radiatively into electron
     and/or photon.  The kinetic decoupling temperature is set to be
     $1$~MeV. The dashed line denotes the canonical annihilation cross
     section ($= 3\times 10^{-26} {\rm cm}^{3}{\rm sec}^{-1}$ ).}
   \label{fig:BBN_rad}
 \end{center}
\end{figure}

Here we discuss the case of an electromagnetic annihilation modes into
electron and/or photon. It is notable that the total amount of
energies into electromagnetic modes approximately determines the
bound, independently of the detail of each mode.  In
Fig.~\ref{fig:BBN_rad} we plot the upper bounds on the annihilation
cross section obtained from the observational limit on $^3$He/D, with
$n=1$ (top) and $n=2$ (bottom) for various values of $\epsilon =
10^{-10}$ -- $10^{-3}$. 
Here the kinetic decoupling temperature is set to be $1$~MeV. The
dashed line denotes the canonical annihilation cross section ($=
3\times 10^{-26} {\rm cm}^{3}/{\rm sec}$ ). In the top panel, we see
that the bounds highly depend on the cutoff parameter $\epsilon$ when
$\epsilon \gtrsim 10^{-7}$.  This behavior can be understood from
  the fact that the production of ${\rm ^3 He}$ becomes most efficient
  when $T\sim 10^{-4}\ {\rm MeV}$; at such a temperature, the
  enhancement factor is estimated as $R_{e}^{-1} \sim 5\times10^{-7}
  \left(T_{\rm kd} /{\rm MeV} \right)^{-1/2} \left(m/{\rm TeV}
  \right)^{-1/2}(T/10^{-4}{\rm MeV})$,
 which becomes smaller than $\sim 10^{-7}$ with the
  present choice of parameters.  Then, when $\epsilon \gtrsim
  10^{-7}$, the cross section is enhanced purely by the factor of
  $\epsilon^{-1}$.  To allow the canonical value of the annihilation
cross section for a few TeV mass of dark matter, we need $\epsilon
\gtrsim 10^{-4.5}$ at least. In the case of $n=2$ which is plotted in
the bottom panel of Fig.~\ref{fig:BBN_rad}, $R_{e}^{-1}$ is much
smaller than $\epsilon$ everywhere in this parameter space. Therefore
$\epsilon^{-1}$ determines the enhancement of the annihilation cross
section, and there exists a simple scaling law for the line of the
limits, which means that the upper bound is proportional to
$\epsilon$.

This feature is slightly different in case of $T_{\rm
  kd}=1$~keV. Because the inverse of the enhancement factor with $n=1$
is the order of $R_{e}^{-1} \sim 1 \times 10^{-5}$, any constraints
with $\epsilon \lesssim 1 \times 10^{-5}$ is insensitive to
$\epsilon$. In case of $n=2$, the constraint is same as the bottom
panel of Fig.~\ref{fig:BBN_rad} because of the same reason.


\begin{figure}
 \begin{center}
   \includegraphics[width=0.6\linewidth]{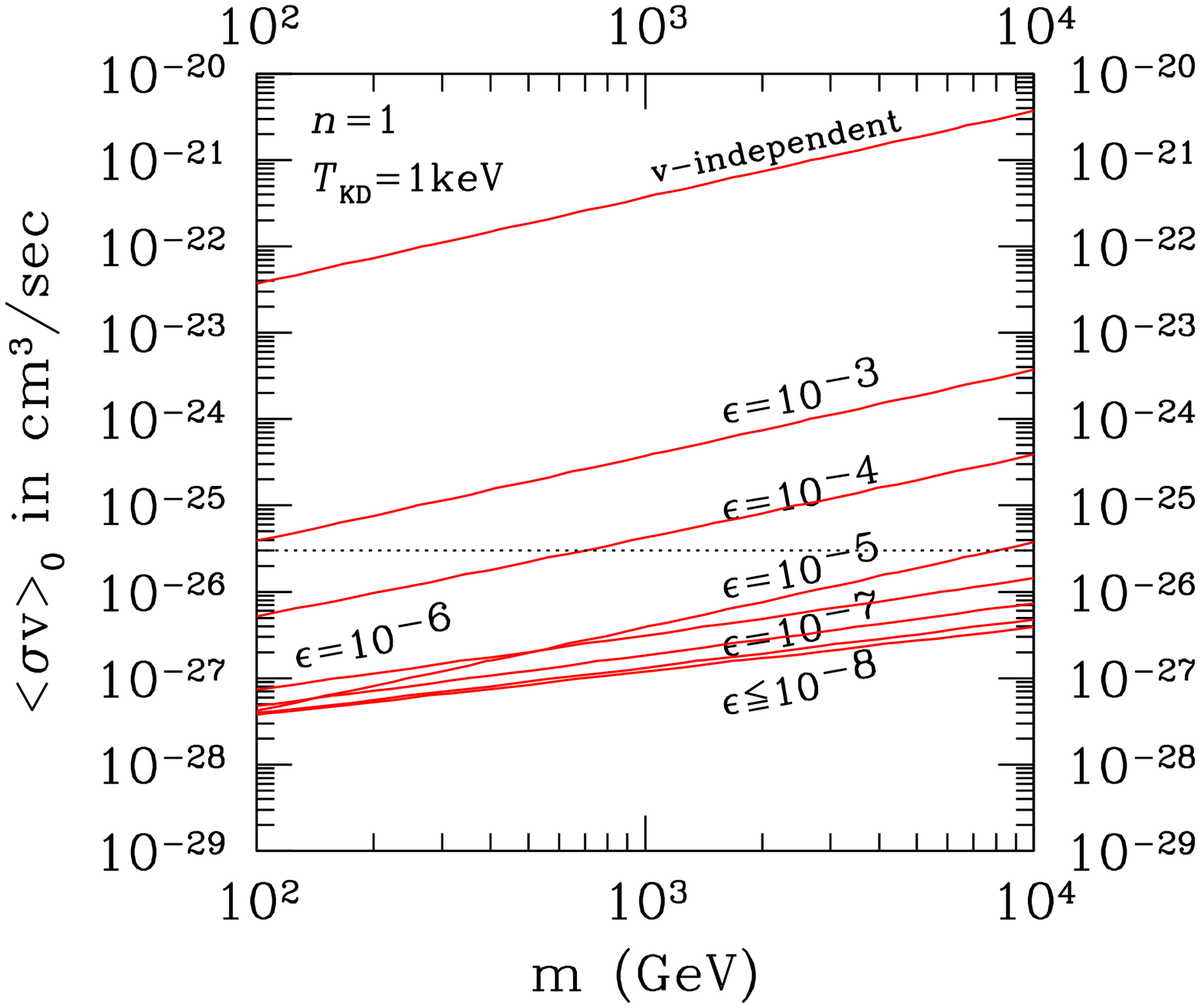}
   \caption{ Same as Fig.~\ref{fig:BBN_rad}, but for the kinetic
   decoupling temperature set to be $1$~keV. The case of $n=2$ is
   completely same as the  bottom panel of Fig.~\ref{fig:BBN_rad}.} 
   \label{fig:BBN_radkeV}
   \vspace{1.0cm}
 \end{center}
\end{figure}

\subsubsection{Constraints on hadron injection} 
\label{subsubsec:hadron}

\begin{figure}
 \begin{center}
   \includegraphics[width=0.6\linewidth]{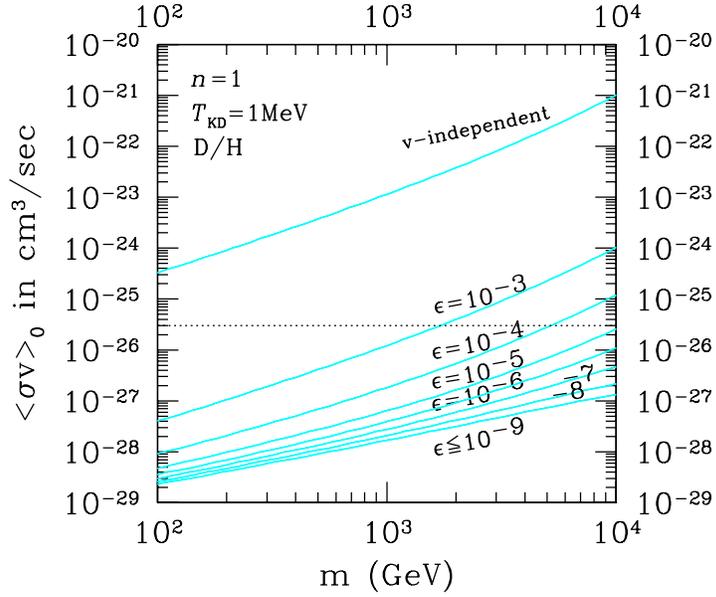}
   \includegraphics[width=0.6\linewidth]{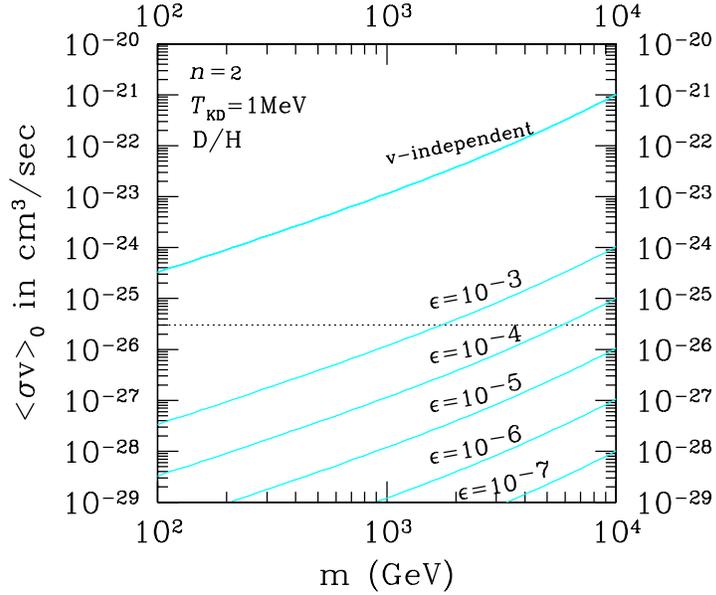}
   \caption{Upper bound on the annihilation cross section obtained
     from the observational D/H limit with $n=1$ (top) and $n=2$
     (bottom) for various values of $\epsilon = 10^{-10}$ --
     $10^{-3}$. Here DM is assumed to annihilate into a $W$-boson
     pair.  The kinetic decoupling temperature is set to be
     $1$~MeV. The dashed line denotes the canonical annihilation cross
     section ($=3\times 10^{-26} {\rm cm}^{3}/{\rm sec}$ ) which
     gives the right amount of the dark-matter relic density.}
   \label{fig:BBN_hadDH}
 \end{center}
\end{figure}

When we consider the injection of hadronic particles,
the limit is completely different from that of the electromagnetic
particles.  The constraint on the overproduction of the deuterium due
to the $^{4}$He destruction often gives the most stringent
  constraint~\cite{Hisano:2009rc}.  To study the hadronic
  injection, hereafter, we assume DM annihilates into a $W$-boson pair
  as a typical hadronic DM annihilation channel; in such a case,
  significant amount of hadrons are produced by the subsequent decay
  of the $W$ bosons produced by the DM annihilation.  Constraints do
not change much for other cases, such as DM annihilation into $b\bar
b$~\cite{Hisano:2009rc}.  In Fig.~\ref{fig:BBN_hadDH} we plot the
upper bound on the annihilation cross section obtained from the
observational limit on D/H with $n=1$ (top) and $n=2$ (bottom). The
kinetic decoupling temperature is set to be $1$~MeV.  First let us
consider the case of $n=1$.  Because the hadrodissociation
  processes become most effective at $T \sim 10^{-2}$ MeV, for which
  the enhancement factor is estimated as $R_{e}^{-1} \sim 5\times
  10^{-5} \left(T_{\rm kd} /{\rm MeV} \right)^{-1/2} \left(m/{\rm TeV}
  \right)^{-1/2}(T/10^{-2}{\rm MeV})$, 
the constraint is determined only by the value of
  $\epsilon$ if $\epsilon \gtrsim 10^{-5}$.  To agree with the
canonical annihilation cross section, we have to assume $\epsilon
\gtrsim 10^{-3}$ for a few TeV mass of dark matter.

\begin{figure}
 \begin{center}
   \includegraphics[width=0.6\linewidth]{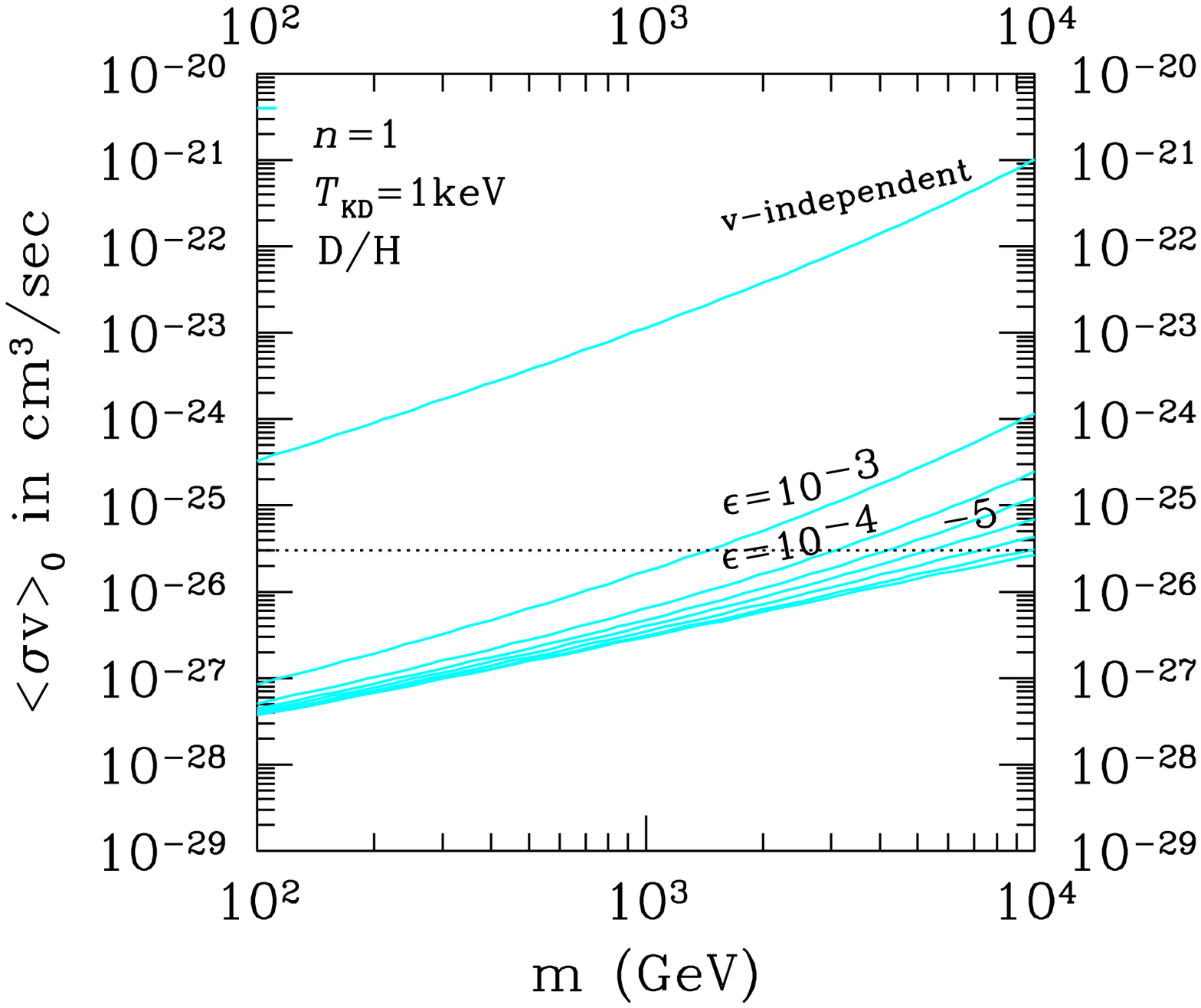}
    \caption{Same as Fig.~\ref{fig:BBN_hadDH}, but for the kinetic
    decoupling temperature set to be $1$~keV. The case of $n=2$ is
     same as the  bottom plot of Fig.~\ref{fig:BBN_hadDH}.}
   \label{fig:BBN_hadDHkeV}
 \end{center}
\end{figure}

If the kinetic decoupling occurs at around 1 keV, the enhancement
factor behaves differently from the case of $T_{\rm kd}=1$ MeV because
the hadrodissociation occurs before the time of the kinetic
decoupling. As is shown in Fig.~\ref{fig:BBN_hadDHkeV}, then the
enhancement factor is estimated to be $R_{e}^{-1} \sim 5 \times
10^{-4} \left(T /10^{-2}{\rm MeV} \right)^{1/2} \left(m/{\rm TeV}
\right)^{-1/2}$ at $T = 10^{-2}$~MeV, from which we easily find that
the constraint is independent of $\epsilon$ for $\epsilon \lesssim
10^{-4}$

When we consider $n=2$, the cutoff parameter determines the upper
bound everywhere in the current parameter space  for both $T_{\rm kd}
= $ 1~MeV and 1~keV. The result is shown in the bottom panel of
Fig.~\ref{fig:BBN_hadDH} as a representative of both cases.

Although so far we have discussed the limit from D/H, consideration of
other light elements sometimes tighten the constraint. In particular
the limit from $^{3}$He/D by the photodissociation of $^{4}$He could
also give us stronger limits. Note that even in the annihilation into
quarks and gluons, the photodissociation occurs because a sizable
amount of the electromagnetic particles is also injected as decay
products.  For example, the electromagnetic energy corresponds to
$\sim 47 \%$ of the total energy in case of the annihilation into a
$W$-boson pair~\cite{Hisano:2009rc}.  In
Refs.~\cite{Hisano:2008ti,Hisano:2009rc,Jedamzik:2009uy}, it has been
shown that the upper bound from $^{3}$He/D due to the
photodissociation accompanied with the hadronic annihilation is much
weaker than that from D/H when the annihilation cross section does not
depend on $v$. The bound on $\epsilon$ from D/H was severer than the
one from $^{3}$He/D by three or four order of magnitude. On the other
hand in case of the $v$-dependent cross section, the situation can be
altered since the enhancement factor depends on the temperature.


\begin{figure}
 \begin{center}
   \includegraphics[width=0.6\linewidth]{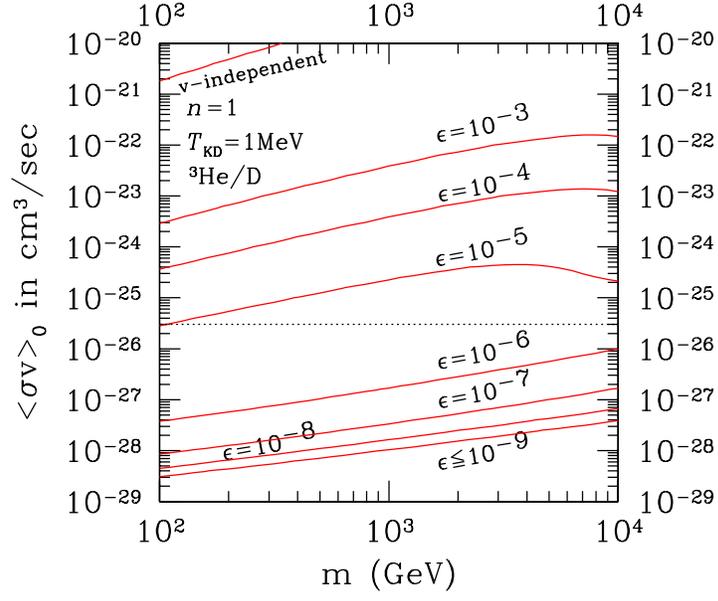}
   \includegraphics[width=0.6\linewidth]{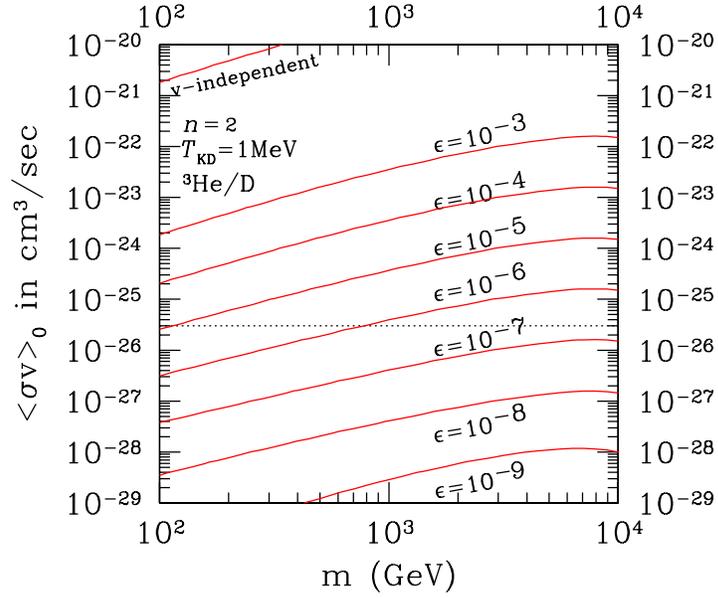}
    \caption{Same as Fig.~\ref{fig:BBN_rad}, but for DM annihilating into
        a $W$-boson pair.}
   \label{fig:BBN_hadHe3D}
 \end{center}
\end{figure}

Because the constraint on $^{3}$He/D is sensitive to the cosmic
history at $T \sim 10^{-4}$ MeV, the enhancement factor is the order
of $\sim 10^{7}$ for $T_{\rm kd}= 1$ MeV with $n=1$. For a small
cutoff parameter $\epsilon \lesssim 10^{-8}$, then the constraint from
$^{3}$He/D can become stronger than that from D/H for $m \gtrsim 1$~TeV. This
feature is shown in the top panel of Fig.~\ref{fig:BBN_hadHe3D}.  In
this figure we plot the constraint only from $^{3}$He/D, ignoring the
D/H constraint. Notice that large amounts of D are produced in most
parameter space as is seen from Fig.~\ref{fig:BBN_hadDH}.  Since
hadro/photo-dissociations of $^4$He also create $^3$He, both D and
$^3$He are produced from the standard BBN and hadro/photo-dissociation
processes.  For sufficiently large $\langle\sigma v\rangle_0$, both D
and $^3$He are  produced by the hadrodissociation of $^4$He
at $T\sim 10^{-2}$MeV, and the constraint comes from the additional
photodissociation effects at around $T\sim 10^{-4}$MeV.  These mixed
processes complicate the behavior of the lines in
Fig.~\ref{fig:BBN_hadHe3D}. There is no simple scaling law among lines
with respect to the line of $v$-independent constraint.
On the other hand, if we take $n=2$ , the bound from $^{3}$He/D is
always weaker than that of D/H. This is clearly shown in the bottom
panel of Fig.~\ref{fig:BBN_hadHe3D}.



\begin{figure}
 \begin{center}
   \includegraphics[width=0.6\linewidth]{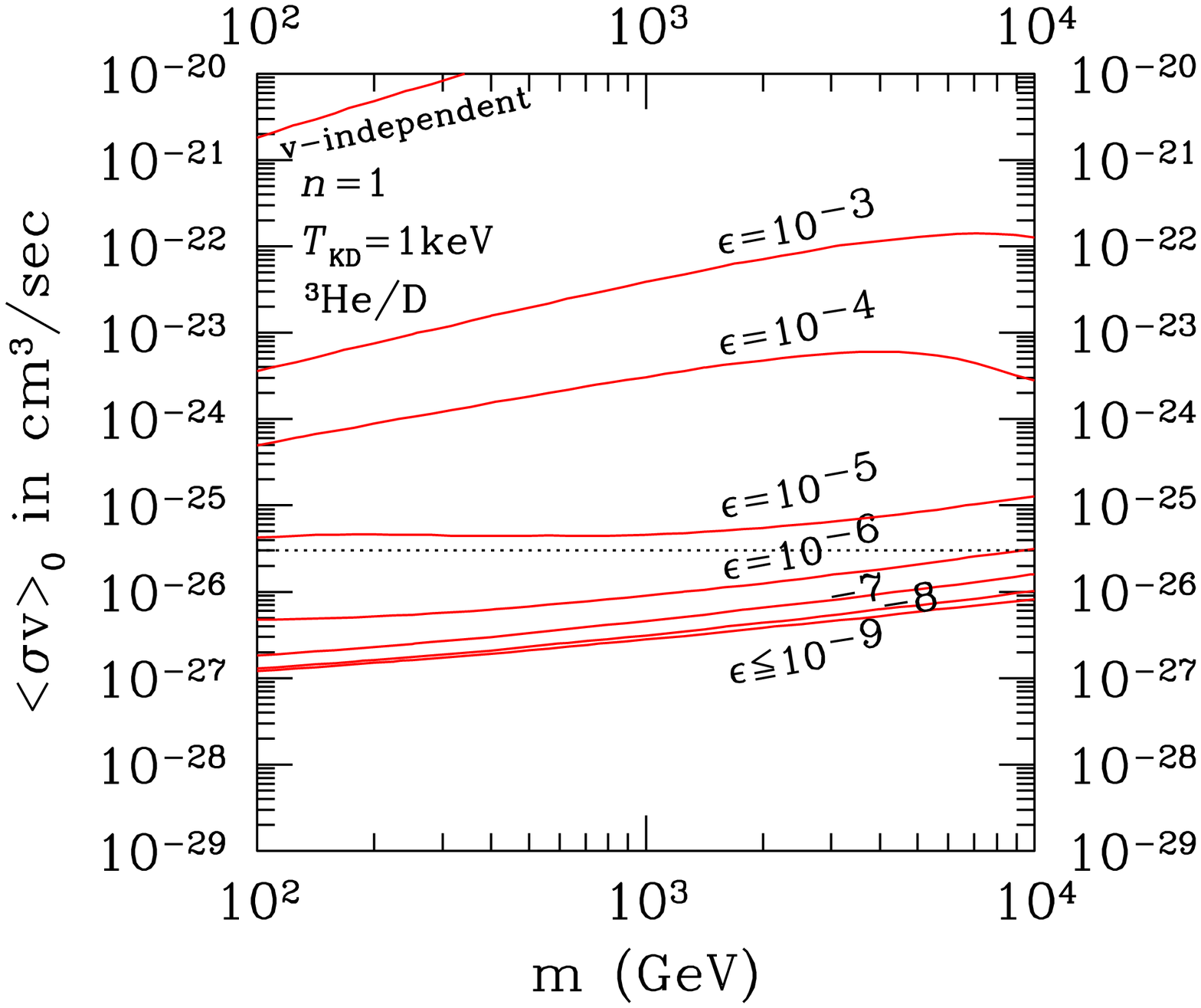}
   \includegraphics[width=0.6\linewidth]{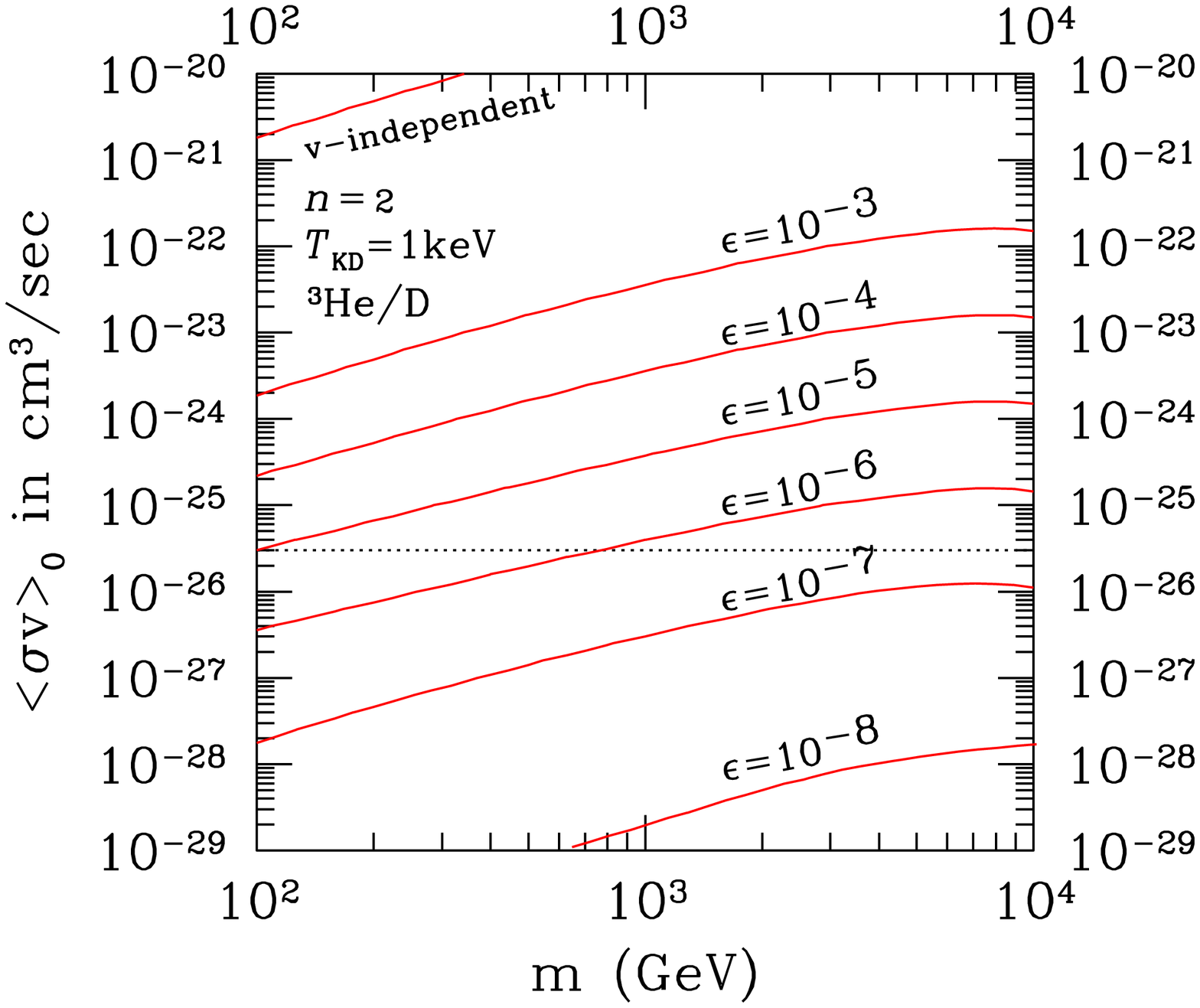}
    \caption{Same as Fig.~\ref{fig:BBN_radkeV}, but for DM annihilating into
    a $W$-boson pair with $n=1$ (top) and $n=2$ (bottom) .}
   \label{fig:BBN_hadHe3DkeV}
 \end{center}
\end{figure}


  We also consider the case of $T_{\rm kd}=1$~keV. Results are shown
  in Fig.~\ref{fig:BBN_hadHe3DkeV}.  As in the previous case, the
  constraint is mostly from the abundance of D.  In some parameter
  region, however, ${\rm ^3 He}$/D gives the most stringent
  constraint. This fact is seen in the case of $n=1$ for $m \gtrsim
  1$~TeV and $\epsilon \lesssim 10^{-6}$.
  In addition, for $n=2$ (the bottom panel of
  Fig.~\ref{fig:BBN_hadHe3DkeV}), a simple scaling law (i.e., the
  proportionality of the upper bound on $\langle\sigma v\rangle_0$ to
  $\epsilon$) breaks down once $\epsilon$ becomes smaller than
  $10^{-8}$; for such a small value of $\epsilon$, the constraint
  becomes significantly stringent.  This feature can be understood
  analytically because, for $\epsilon \lesssim 10^{-8}$, $\Delta
  \rho/s$ starts to increase as a function of $t$ after $t=10^{6}$~sec
  ($T=1$~keV). Such a behavior is clearly seen as the dashed line in
  the bottom panel of Fig.~\ref{fig:rhos}.

Before closing this subsection, we comment on the constraints from the
Li abundances.  Photo/hadro-dissociation processes also modify
abundances of $^6$Li and $^7$Li.  However, constraints from
observations of $^7$Li/H and $^6$Li/H are weaker than those from D/H
and/or $^3$He/D for annihilating DM~\cite{Hisano:2009rc}.

\subsection{Constraints from CMB} \label{sec:CMB}

Energy injection around the recombination epoch affects the CMB
anisotropy~\cite{Chen:2003gz,Padmanabhan:2005es,Galli:2009zc,Kanzaki:2009hf}.\footnote{
  DM annihilation also induces CMB spectral distortion, which is
  constrained from COBE FIRAS measurement~\cite{Fixsen:1996nj}.  This
  constraint is weaker than that from anisotropy
  measurements~\cite{Zavala:2009mi,Hannestad:2010zt}.  } This is
because injected energy can ionize neutral hydrogens and modify the
standard recombination history of the Universe.  The effect is
characterized by the quantity $d\chi_{\rm ion}^{(i)}(E,z',z)$, which
represents the fraction of injected electron (photon) energy $E$ for
$i=e (\gamma)$ at the redshift $z'$ used for ionization of the
hydrogen atom at the redshift between $z$ and $z+dz$.  The evolution
equation of the ionization fraction of the hydrogen atom, $x_e$,
includes the following additional term,
\begin{equation}
        -\left[ \frac{dx_e}{dz} \right]_{\rm DM} = 
        \int \frac{dz'}{H(z')(1+z')}\frac{n_{\rm DM}
        ^2(z') \langle \sigma v\rangle}{n_H(z')}
        \frac{m}{E_{\rm Ry}}\frac{d\chi_{\rm ion}^{(F)}(m,z',z)}{dz},
\end{equation}
where $E_{\rm Ry}=13.6~$eV is the Rydberg energy, $n_H$ is the number
density of the hydrogen atom and
\begin{equation}
  \frac{d\chi_{\rm ion}^{(F)}(m,z',z)}{dz}=\int dE \frac{E}{m} \left[ 
    \frac{dN_e^{(F)}}{dE}\frac{d\chi_{\rm ion}^{(e)}(E,z',z)}{dz}
    +\frac{1}{2}\frac{dN_\gamma^{(F)}}{dE}\frac{d\chi_{\rm ion}^{(\gamma)}(E,z',z)}{dz}
  \right ].
\end{equation}
Here $F$ denotes the final state of the DM annihilation, e.g.,
$F=e^+e^-$, $W^+W^-$, etc., and $ dN_{e,\gamma}^{(F)}/dE $ denotes the
energy spectrum of the electron (photon) generated from the cascade
decay of $F$.  In the case of $F = e^+e^-$, we have $dN_{e}^{(F)}/dE =
\delta(E-m)$.  For general final state $F$, it is evaluated by the
PYTHIA code~\cite{Sjostrand:2006za}.  We follow the methods described
in Refs.~\cite{Kanzaki:2009hf,Kanzaki:2008qb} to compute $d\chi_{\rm
  ion}(E,z',z)/dz$.  This term is included in the RECFAST
code~\cite{Seager:1999bc} implemented in the CAMB
code~\cite{Lewis:1999bs} for calculating the CMB anisotropy.


\begin{figure}
  \begin{center}
    \includegraphics[width=0.6\linewidth]{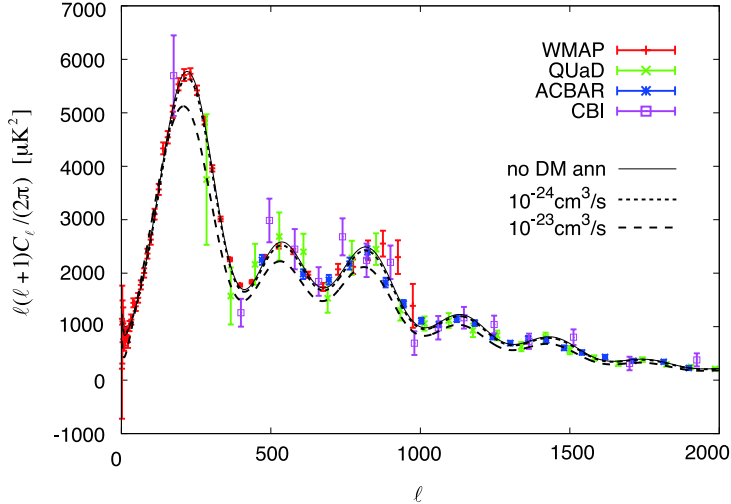}
    \caption{ Power spectrum of the CMB anisotropy with no DM
      annihilation effect (solid), with $\langle \sigma v \rangle =
      10^{-24}{\rm cm^3/sec}$ (dotted) and $\langle \sigma v \rangle =
      10^{-23}{\rm cm^3/sec}$ (dashed) for $m=1$TeV and assuming DM
      annihilation into $e^+e^-$ with velocity-independent
      annihilation cross section.  Also shown are data points from
      WMAP, QUaD, ACBAR and CBI.  }
    \label{fig:Cl}
  \end{center}
\end{figure}


Additional energy injections from DM annihilation around the
recombination epoch cause ionization of neutral hydrogen atoms.
  Thus the effect is to slow down the recombination of the Universe.
  As a result, anisotropies in CMB are dumped at small scales due to
  the increase in thickness of the last scattering surface.
Fig.~\ref{fig:Cl} shows the $TT$ power spectrum of the CMB temperature
anisotropy, with/without DM annihilation effect.  The solid line
corresponds to the best-fit $\Lambda$CDM model without DM
annihilation, and dotted line to DM annihilation cross section into
$e^+e^-$ with $\langle \sigma v \rangle = 10^{-24}{\rm cm^3/sec}$ and
dashed line to DM annihilation cross section with $\langle \sigma v
\rangle = 10^{-23} {\rm cm^3/sec}$, while all cosmological parameters
are fixed.  Here we have taken $m=1$TeV with velocity-independent
annihilation cross section.  It is seen that DM annihilation effects
suppress the $TT$ spectrum, reflecting the increase in thickness of the
last scattering surface.

It is not hard to imagine that this effect has a degeneracy with other
cosmological parameters.  In particular, the increase of the
reionization optical depth causes similar effects.  In order to derive
conservative bounds on the DM annihilation cross section, we must take
into account degeneracies between DM annihilation effect and other
cosmological parameters.  We have derived 2$\sigma$ constraints
  using a profile likelihood function where the other cosmological
  parameters including the six standard ones 
  ($\omega_b, \omega_c,\Omega_\Lambda,n_s,\tau,\Delta_{\mathcal R}^2$ in the notation of Ref.~\cite{Dunkley:2008ie}) 
  and the amplitude of the
  Sunyaev-Zel'dovich effect are marginalized so that the original
  likelihood function is maximized for given DM annihilation cross
  section and mass.  The likelihood surface is scanned by using the
  CosmoMC code~\cite{Lewis:2002ah}; in our analysis, we have modified
  the CosmoMC code to take account of the above mentioned effect of
  energy injection. The used datasets include
WMAP~\cite{Dunkley:2008ie}, ACBAR~\cite{Reichardt:2008ay},
CBI~\cite{Sievers:2009ah} and QUaD~\cite{Pryke:2008xp}.  
As opposed to BBN constraints, CMB constraint depends on the injected
radiative energy, hence purely leptonic annihilation is more strongly
constrained than the hadronic one.

\begin{figure}
  \begin{center}
    \includegraphics[width=0.6\linewidth]{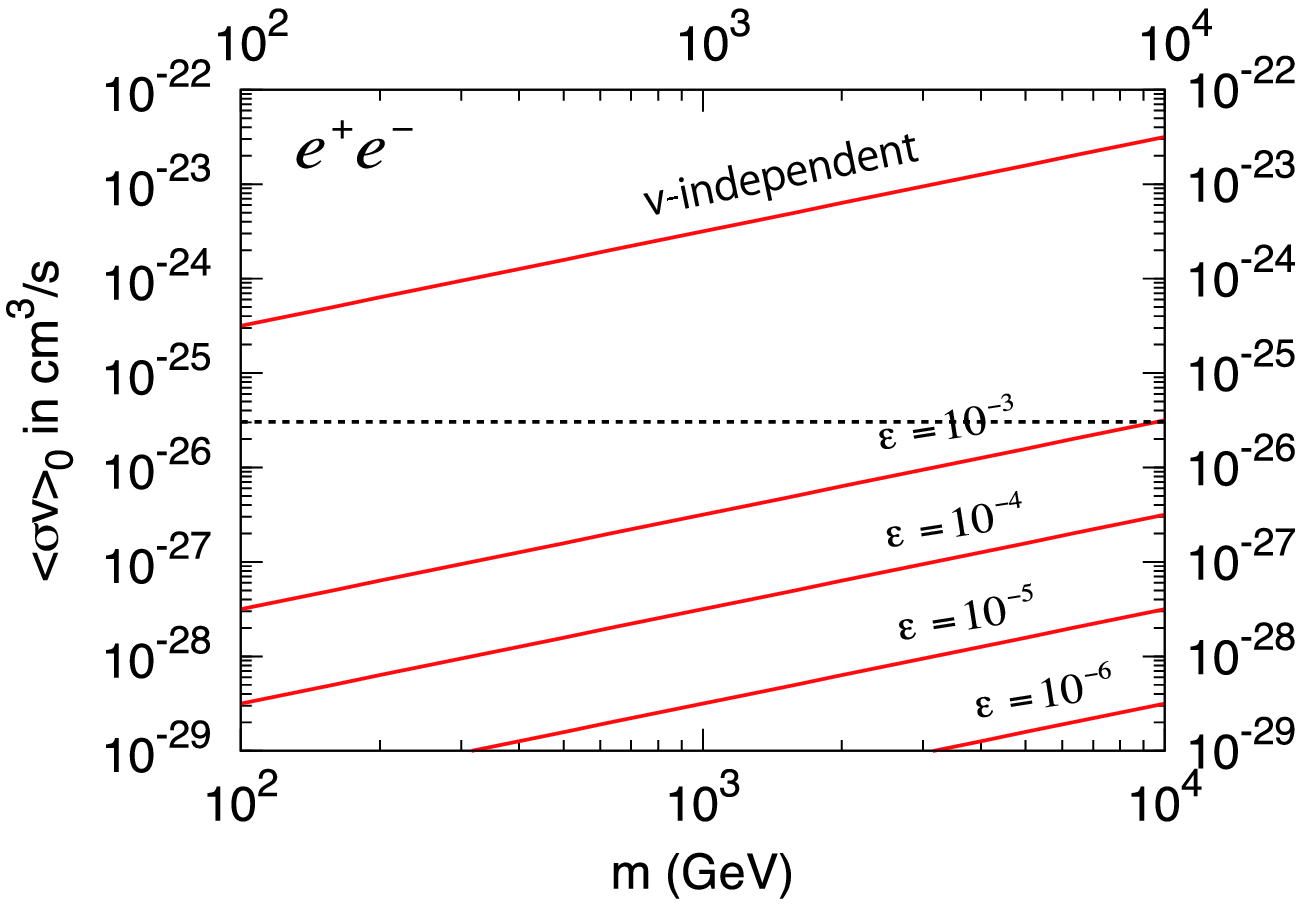}
    \vskip 1cm
    \includegraphics[width=0.6\linewidth]{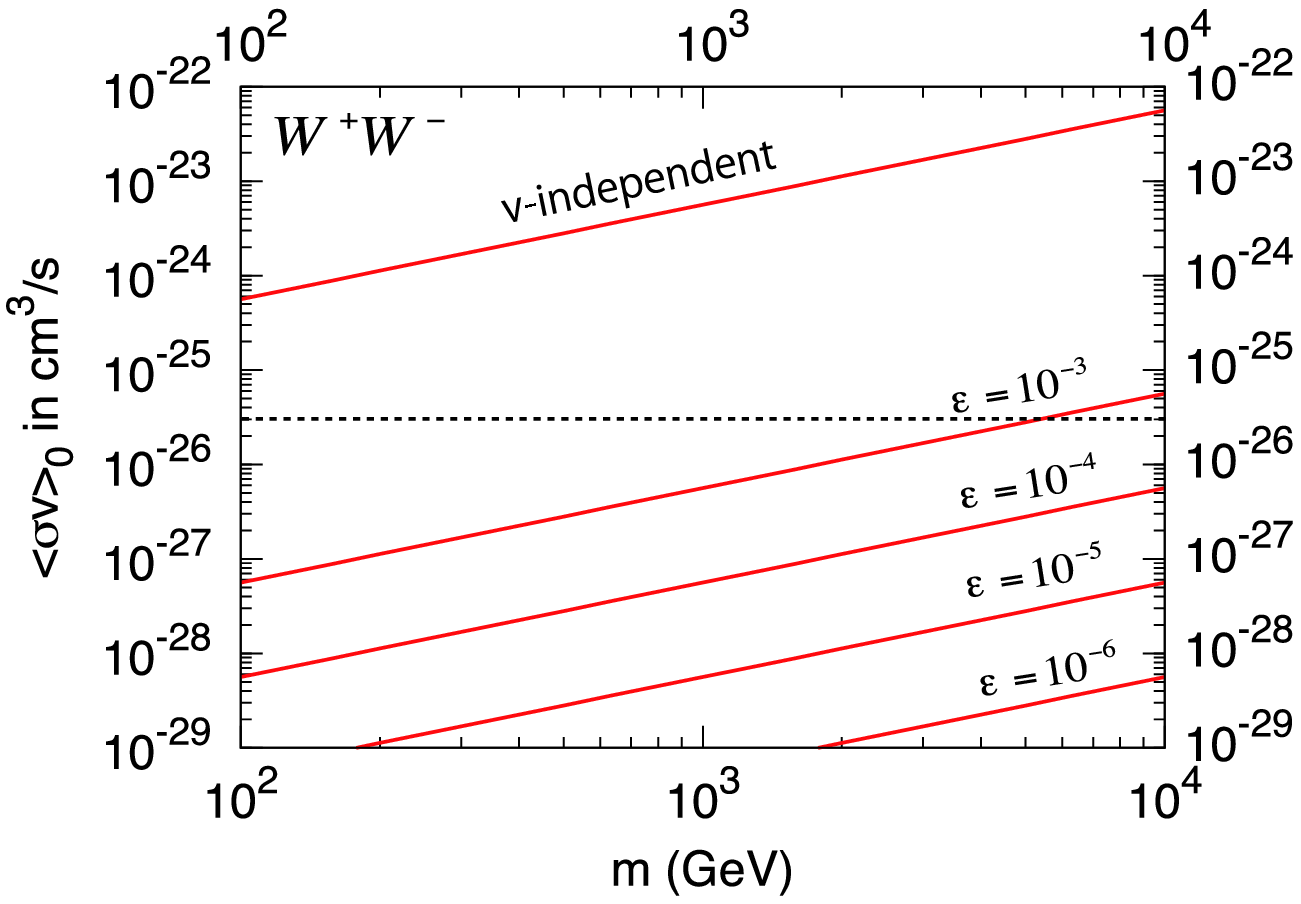}
    \caption{Upper bound on the annihilation cross section obtained
      from CMB anisotropy data as a function of DM mass for $\epsilon
      = 10^{-3}$ -- $10^{-7}$.  DM is assumed to annihilate into
      $e^+e^-$ pair in the top panel and $W^+W^-$ in the bottom panel.
      Here we have taken $n=1$ and $T_{\rm kd} = 1~{\rm MeV}$.
      Results do not change for $n=2$ and/or $T_{\rm kd} = 1~{\rm
        keV}$. }
    \label{fig:CMB}
  \end{center}
\end{figure}

The result is presented in Fig.~\ref{fig:CMB} where we plot upper
bounds on the annihilation cross section obtained from CMB anisotropy
data as a function of DM mass for $\epsilon = 10^{-3}$ -- $10^{-7}$.
DM is assumed to annihilate into $e^+e^-$ pair in the top panel and
$W^+W^-$ in the bottom panel.  Here we have taken $n=1$ and $T_{\rm
  kd} = 1~{\rm MeV}$.  We have checked that the results do not change
for $n=2$ and/or $T_{\rm kd} = 1~{\rm keV}$.  This is because the CMB
constraint is sensitive to the annihilation rate at around the
recombination epoch, $T \lesssim 1$~eV, and hence the annihilation
cross section is already saturated for most interesting range of
$\epsilon$ for both $n=1$ and $n=2$.  Comparing them with BBN
constraints, it is found that the CMB constraint is severer for the
leptonic annihilation case independently of the parameters.

In the case of hadronic annihilation with $m \lesssim$~a few TeV, the
situation is not so simple.  For $n=1$ and $T_{\rm kd}=1$MeV, CMB
gives weaker constraint than BBN for $\epsilon \gtrsim 10^{-4}$, as
seen from Fig.~\ref{fig:BBN_hadDH}, but becomes tighter for $\epsilon
\lesssim 10^{-4}$.  The situation is similar for $T_{\rm kd}=1$keV.
This is because the BBN constraint from the observation of D/H is
sensitive to the annihilation at $T\sim 10^{-2}$MeV, and the
annihilation cross section do not saturate at that epoch for small
$\epsilon$ for $n=1$.  On the other hand, for $n=2$, BBN gives tighter
constraint than the CMB for parameter ranges shown in the figures.

Therefore, CMB takes complementary role to BBN in constraining the DM
annihilation with velocity-dependent annihilation cross section.


\section{Conclusions} \label{sec:conc}

In this paper we have investigated effects of DM annihilation on BBN
and CMB, and derived constraints on the DM annihilation rate,
particularly focusing on the case where the annihilation cross section
has a velocity-dependent structure.  This is partly motivated by the
observations of cosmic-ray positron/electron excesses and their
explanations by the DM annihilation contribution.  We
phenomenologically parametrized the velocity-dependence of the
annihilation cross section and the critical velocity at which such an
enhancement saturates, and derived general constraints on them.  Our
constraints are applicable to known velocity-dependent DM annihilation
models, such as the Sommerfeld and Breit-Wigner enhancement scenarios.

These results have been plotted in
Figs.~\ref{fig:BBN_hadDH}~--~\ref{fig:CMB} by changing the parameters
and observation. Therefore readers can read off the allowed parameter
regions from those figures in accordance with the intended use.

Some comments are in order.  If the DM annihilation is
helicity-suppressed, the $p$-wave process may be the dominant mode, as
is often the case with Majorana fermion DM.  In this case we obtain
$n=-2$ : $\langle \sigma v \rangle \propto v^2$.  Thus the
annihilation cross section becomes smaller as the temperature
decreases, until the $S$-wave process becomes efficient.  For negative
$n$, the BBN/CMB constraints are weaker than the velocity-independent case.

In the Sommerfeld enhancement scenario, it was pointed out that the
DM-DM scattering mediated by light particle exchanges causes
observationally relevant effect~\cite{Feng:2009hw}, and this also
gives significant constraint~\cite{Feng:2010zp,Finkbeiner:2010sm}.


\section*{Acknowledgment}

This work is supported by Grant-in-Aid for Scientific research from
the Ministry of Education, Science, Sports, and Culture (MEXT), Japan,
No.\ 20244037 (J.H.), No.\ 20540252 (J.H.), No.\ 22244021 (J.H. and
T.M.), No.\ 14102004 (M.K.), No.\ 21111006 (M.K., K.K. and K.N.), No.\
22244030 (K.K. and K.N.), No.\ 18071001 (K.K.), and No.\ 22540263
(T.M.), and also by World Premier International Research Center
Initiative (WPI Initiative), MEXT, Japan.  K.K. was also partly
supported by the Center for the Promotion of Integrated Sciences
(CPIS) of Sokendai.


{}

\end{document}